\documentstyle[pra,aps,multicol,psfig]{revtex}

\begin{document}


\draft

\title{Cooper pair formation in trapped atomic Fermi gases}

\author{M. Houbiers and H. T. C. Stoof}
\address{Institute for Theoretical Physics, University of Utrecht,
         Princetonplein 5, 3584 CC Utrecht, The Netherlands}
        
\maketitle

\begin{abstract}
We apply the closed time-path formalism to evaluate the
dynamics of the BCS transition to the superfluid state
in trapped atomic $^6$Li. We find that the Fokker-Planck equation
for the probability distribution of the order 
parameter is, sufficiently close to the critical temperature,
identical to the equation that describes the switching on of a 
single-mode laser.   
\end{abstract}


\begin{multicols}{2}

After the successful experiments to trap and cool atomic gases 
of the bosonic alkalies $^{87}$Rb, $^7$Li and $^{23}$Na 
\cite{JILA,rice,MIT} below the critical temperature for 
Bose-Einstein condensation, the field of atomic physics has attracted
a lot of interest. Besides the ongoing experiments on the properties 
of the Bose condensate in these gases, 
several groups have started experiments in order to observe also quantum
degeneracy effects in fermionic gases. A particularly promising 
candidate in this respect is the fermionic isotope $^6$Li which is predicted
to make a BCS transition to a superfluid state. Indeed, 
the triplet $s$-wave scattering length $a_s$ of this atom is enormously 
large and negative \cite{randy2}, resulting in a critical temperature of
the same order of those of the BEC experiments \cite{us}. A requirement
for this relatively high critical temperature is that a {\it mixture} of
two different hyperfine levels of the $^6$Li atom must be trapped. 
Depending on the precise hyperfine states trapped, this
can be done in a magnetic or optical trap.
In either case, a study of the dynamics of this BCS phase transition
offers the exciting opportunity to observe the evolution of a spontaneous 
breaking of symmetry under almost ideal conditions.

In case of a magnetically trapped gas mixture, a drawback of the large 
$s$-wave scattering length is the corresponding large decay rate
and resulting limited lifetime of the gas
\cite{us,ourCC}. 
Fortunately, this decay can be suppressed by applying a magnetic 
bias field. For example, the lifetime of a gas with a typical 
density of $2 \times 10^{12}$ cm$^{-3}$ atoms per hyperfine state, 
is of the order of 1 second in
a magnetic bias field of about 7 T, whereas it is of the order of only 0.2 ms
at 0.1 T. Evidently, the lifetime of the gas must be longer than the 
time scale on which the formation of Cooper pairs occurs.
In previous work \cite{us}, we estimated on basis of mean-field theory
the nucleation time to be $O(\hbar/k_B T_c)$. For a homogeneous
mixture with a density of $2\times 10^{12}$ cm$^{-3}$ in each hyperfine
level, the critical temperature was calculated to be 37 nK, resulting
in a nucleation time of $0.2$ ms. Since low magnetic fields are 
experimentally much more convenient, it is also from this point of
view sensible to make a
precise calculation of the nucleation time of the BCS transition.
The aim of the present paper is therefore to evaluate the dynamics of
the BCS phase transition, from which, as a byproduct, 
the time scale for the formation of Cooper pairs can be extracted. 

The dynamics of the phase transition can be described by using the closed
time-path formalism \cite{schwinger}. 
This can easily be understood from an analogy in 
quantum mechanics where the probability amplitude for a particle
moving from a position ${\bf x}_0$ at time $t_0$ to another
position ${\bf x}$ at time $t$ is given by   
$\langle {\bf x},t|{\bf x}_0,t_0\rangle = 
\int {\cal D}[{\bf x}] \exp{ (\imath {\cal A}[{\bf x}]/\hbar) }$ with
${\cal A}[{\bf x}] = \int_{t_0}^t dt' {\cal L}[{\bf x},\dot{{\bf x}}]$ 
the action and ${\cal L}$ the lagrangian of the particle.
The probability distribution for being at position ${\bf x}$ at time $t$, 
i.e., $P[{\bf x},t] =  \int d {\bf x}_0 \langle {\bf x},t|{\bf x}_0,t_0\rangle  
\langle {\bf x}_0,t_0|{\bf x},t\rangle P[{\bf x}_0, t_0] $, is now
given by a path integral over the periodic paths $[{\bf x}(t')]$ 
where $t'$ runs from $t$ to $t_0$ (antichronological, lower branch)
and back from $t_0$ to $t$ (chronological, upper branch), 
i.e., over a closed time or Keldysh contour \cite{schwinger,keldysh}.

In our case, we want to study the dynamics of the BCS phase transition.
Analogous to the simple quantum mechanical description
for the position probability of a particle, we thus need to
derive the time dependence of the probability 
distribution for the order parameter $\Delta$ which describes the superfluid phase
\cite{abrahams,kl,stoof}. The starting 
point is the probability $ P[\psi,\psi^{\ast},t]$ 
for a gas of atoms in two hyperfine states denoted
by $|\alpha\rangle = |\uparrow\rangle, |\downarrow\rangle$  
\[
P[\psi,\psi^{\ast},t] = \int {\cal D}[\psi] {\cal D}[\psi^{\ast}]
\exp{\left( \frac{\imath}{\hbar} {\cal A}[ \psi, \psi^{\ast}] \right)} 
\]
where the functional integral is over the 
fermionic fields $\psi_{\alpha}({\bf x},t)$ and $\psi_{\alpha}^{\ast}
({\bf x},t)$ that are defined on the Keldysh contour $C$ with
$t_0 \rightarrow - \infty$. 
The action ${\cal A}[\psi,\psi^{\ast}]$ is given by
\begin{eqnarray*}
{\cal A}[\psi,\psi^{\ast}] & = &\sum_{\alpha} \int_{C} dt \int d{\bf x}
\psi_{\alpha}^{\ast} ({\bf x},t) \left( \imath \hbar \frac{\partial}{
\partial t} + \frac{\hbar^2 {\bf \nabla}^2}{2 m} + \mu_{\alpha} \right. \\
& & - \left. \frac{V_0}{2}  \psi_{-\alpha}^{\ast}({\bf x},t)  
\psi_{-\alpha}({\bf x},t) \right) \psi_{\alpha}({\bf x},t), 
\end{eqnarray*} 
where $V_0 <0$ represents a local attractive interatomic interaction.
Introducing by means of a Hubbard-Stratonovich transformation
the order parameter $\Delta({\bf x},t)$, 
the action becomes quadratic in the fermionic fields 
and $\psi$ and $\psi^{\ast}$ can easily be integrated out \cite{kl}.
We thus arrive at an effective field theory for this order parameter 
$\Delta({\bf x},t)$ that equals
\begin{equation}
{\cal A}[\Delta,\Delta^{\ast}] = -\imath \hbar \mbox{Tr}[\ln{G^{-1}}]
+ \int_{C} dt \int d{\bf x} \frac{| \Delta({\bf x},t)|^2}{V_0},
\label{actie}
\end{equation}
where the trace runs over the spin, time and spatial variables,
and the Green's function matrix is defined as
$ G({\bf x},t;{\bf x}',t') \equiv -\imath \langle T_{C} [{\bf \psi}
({\bf x},t) {\bf \psi}^{\ast} ({\bf x}',t')] \rangle $.
The spinor ${\bf \psi}^{\ast}({\bf x},t)$ is 
given by $ {\bf \psi}^{\ast}({\bf x},t) \equiv \left( 
\psi_{\downarrow}^{\ast} ({\bf x},t) \mbox{ , } \psi_{\uparrow} ({\bf x},t) 
\right) $ and $T_{C}$ is the time ordening operator on the 
Keldysh contour $C$, 
taking into account the anti-commuting nature of the fields.

Around the critical temperature $T_c$, the order parameter $\Delta$ is small 
compared to $k_B T$, and Eq.~(\ref{actie}) can be expanded in powers of $\Delta$.
To include to lowest order all relevant physics, 
it suffices to take only the terms proportional to $|\Delta|^2$ and $|\Delta|^4$
into account. 
To calculate these quadratic and quartic parts of the effective action 
for $\Delta$ explicitly, we must remember that the time integrals
within these expressions are over the Keldysh contour, and must be 
rewritten as time integrals over the real axis $-\infty < t' < t$ only.
The Green's function $G_0({\bf x},t;{\bf x}',t')$ defined on the contour,
must then be decomposed into retarded, advanced and Keldysh components.
Furthermore, the order 
parameter $\Delta({\bf x},t)$ on the contour can be decomposed
in a classical part $\phi$ and fluctuations $\xi$ defined on the real time axis 
as 
\[
\Delta ({\bf x},t_{\pm}) = \phi({\bf x},t) \pm \frac{1}{2} \xi({\bf x},t)
\]
where the upper and lower sign refer to the upper and lower branch of
the Keldysh contour, respectively. The important point is that
the fluctuations allow the classical part of the
order parameter to evolve from zero to a nonzero value when the system is 
driven through the phase transition, and we can thus indeed describe the 
dynamics of the classical field $\phi$ around
the critical temperature by means of this closed time-path method. 
In contrast, this cannot be achieved within the framework of mean-field theory,
where the order parameter always remains zero if it is zero initially.

Applying subsequently a gradient expansion which assumes that
spatial and temporal variations in the order parameter are small
on the size of the Cooper pairs,
the quadratic plus quartic part of the effective action Eq.~(\ref{actie}) 
up to quadratic order in the fluctuations, becomes
\begin{eqnarray}
{\cal A}[\phi,\phi^{\ast},\xi,\xi^{\ast}]  & = & \left. \int d {\bf x} 
\int_{-\infty}^{t} dt' \right\{  
\phi({\bf x},t') L^{(R)}({\bf x},t') 
\xi^{\ast}({\bf x},t') \nonumber \\ 
 & + &  \xi({\bf x},t') L^{(A)}({\bf x},t') 
\phi^{\ast}({\bf x},t')  \nonumber \\
& + &  \left. \frac{1}{2} \xi({\bf x},t') L^{(K)}({\bf x}, t') 
\xi^{\ast}({\bf x},t') \right\},
\label{actie2FT}
\end{eqnarray}
where to lowest non-vanishing 
order in ${\bf \nabla}$ and in $\partial/\partial
t$ the retarded and advanced factors are \cite{abrahams,kl}
\begin{equation}
L^{(R,A)} ({\bf x},t)  =  \alpha + \gamma {\bf \nabla}^2 
- \beta |\phi({\bf x},t)|^2 \pm N_0 \frac{\pi}{8} \frac{\hbar}{k_B T}
\frac{\partial}{\partial t} 
\label{LRA}
\end{equation}
and the Keldysh part is given by
\begin{equation}
L^{(K)} ({\bf x},t)  =   \imath N_0 \frac{\pi}{2}.
\label{LK}
\end{equation}
Here we considered only the optimal case $\mu_{\uparrow} = \mu_{\downarrow}$
and used particle-hole symmetry around the Fermi level of the system.
The constants $\alpha$, $\beta$ and $\gamma$ are given by
\begin{eqnarray}
\alpha  & = & N_0 \left( 1 - \frac{T}{T_c} \right) \label{alpha} \\ 
\beta & = &N_0 \frac{7}{8} \frac{\zeta(3)}{(\pi k_B T)^2} \\ 
\gamma & = & N_0 \frac{7}{8} \frac{\zeta(3)}{6 \pi^2} \left(
\frac{\hbar v_F}{k_BT} \right)^2. \label{gamma} 
\end{eqnarray}
Furthermore, $N_0 = m k_F /(2 \pi^2 \hbar^2)$ 
denotes the density of states at the Fermi level,
the critical temperature obeys 
$k_B T_c \simeq 5/3 \epsilon_F \exp{(-1/\lambda)}$
\cite{us} with $\lambda = 2 k_F |a_s|/\pi$, and 
$v_F$ is the Fermi velocity of the system. 

The last term in the right-hand side of Eq.~(\ref{actie2FT}) describes
the quadratic fluctuations. If one would neglect them, and integrate out the
(in that case linear) fluctuations, one would obtain a (dissipative) non-linear
Schr\"odinger equation for the classical
part $\phi$ of the order parameter which would also arise from
equilibrium theory. Of course, for our non-equilibrium purposes, the
quadratic fluctuations are important. 
The Fourier transforms of Eqs.~(\ref{LRA}) and (\ref{LK}), 
$ L^{(R,A)}({\bf k},\omega)$ and $L^{(K)}({\bf k},
\omega)$ in principle are related through the fluctuation-dissipation theorem  
\begin{equation}
L^{(K)} ({\bf k},\omega)  =  \pm 2 \imath \left(2 N^B (\hbar \omega) + 1
\right) \mbox{Im}[L^{(R),(A)} ({\bf k},\omega)]
\nonumber
\end{equation}
where $N^B (\hbar \omega) $ denotes the Bose distribution evaluated
at $\hbar \omega  $. At low frequencies it thus gives 
$L^{(K)}({\bf k},\omega) =
\pm 2 \imath (2k_BT /\hbar \omega) \mbox{Im}[L^{(R),(A)} ({\bf k},\omega)]$,
in agreement with the above results.
Notice that if we expand the quadratic part of the action ${\cal A}$ in
Eq.~(\ref{actie}) up to lowest non-vanishing order in ${\bf \nabla}$ and
$\partial/\partial t$, 
we only have to expand the quartic part of ${\cal A}$ to zeroth order, 
i.e.~we only need to consider the local contributions proportional to 
$\phi({\bf x},t) \xi^{\ast}({\bf x},t) |\phi({\bf x},t)|^2$ and 
it's complex conjugate. Since the factor $\beta$ has no imaginary part,
we conclude that it is indeed consistent to neglect Keldysh-like terms that
would in principle also arise from the quartic part of 
the action ${\cal A}$. 
   
In order to arrive at an equation of motion for the probability
distribution
of the classical value $\phi$, we first integrate out the fluctuations 
$\xi$ from the action. Before doing so, we however need to realize that 
in real experiments, the gas will be trapped in an
external potential $V_{\mbox{ext}}(x)$ generally in the shape of a 3D isotropic
harmonic oscillator $V(x) = 1/2 \: m \omega_0^2 {\bf x}^2 $.
As a result, the coefficients $\alpha$, $\beta$ and $\gamma$ in 
Eqs.~(\ref{alpha}) to (\ref{gamma}) now depend on ${\bf x}$, since
the Fermi energy $\mu_{\alpha} ({\bf x}) = \mu_{\alpha} - V({\bf x})
- (4 \pi a_s \hbar^2/m) n_{-\alpha}({\bf x})$ and therefore the
densities $n_{\alpha}({\bf x})$ and the critical temperature $T_c({\bf x})$
are position dependent. Hence, we can make use of the following expansion 
in the effective action Eq.~(\ref{actie2FT})
$\phi({\bf x},t)  = \sum_n a_n(t) \phi_n({\bf x})$ and  
$\xi({\bf x},t)  =  \sum_n b_n(t) \phi_n({\bf x})$,
where the normalized wave functions $\phi_n({\bf x})$ are solutions to
Schr\"odinger's equation \cite{baranov}
\begin{equation}
-[\alpha({\bf x})  + \gamma({\bf x}) {\bf \nabla}^2] \phi_n({\bf x})
= N_0({\bf x}) \varepsilon_n \phi_n({\bf x}).
\label{SV1}
\end{equation}
It is well-known 
that close to $T_c$, the spatial region where the order parameter
becomes non-zero, is small compared to the region over which the density
changes \cite{us,baranov}. So, we can take $N_0({\bf x}) \simeq N_0({\bf 0})$,
$v_F({\bf x}) \simeq v_{F}({\bf 0})$, but we must expand $T_c({\bf x})$
around ${\bf x}= {\bf 0}$, since it depends exponentially on the
coupling constant $\lambda({\bf x})$.     

Considering again only 
the optimal case that $\mu_{\downarrow} = \mu_{\uparrow}$ and using
that the density in each hyperfine level around ${\bf x}={\bf 0}$ can be 
approximated with $n_{\alpha}({\bf x}) = 1/2\:n_t({\bf 0}) (1 -  ({\bf x}/R)^2)$,
where $n_t({\bf 0})$ is the total gas density in the center of the trap,
one finds from the zero temperature expression for the respective
chemical potentials that 
\[
R^2 = l^4 \left( 4 \pi n_t({\bf 0}) a + 2 \left( \frac{
\pi^2}{\sqrt{3}} \right)^{2/3} [n_t({\bf 0})]^{2/3} \right), 
\]
where $l=\sqrt{\hbar/m\omega_0}$.
From this expression one can immediately see that the effect of the mean-field
interaction in the gas is to contract the gas cloud to the center of the trap.
Expanding $T_c({\bf x})$ around ${\bf x}={\bf 0}$, one arrives at
\cite{baranov}
\[
\alpha({\bf x}) = N_0({\bf 0}) \left\{ \ln{ \left( \frac{T_c({\bf 0})}{T}
\frac{\epsilon_F({\bf x})}{\epsilon_F({\bf 0})} \right) } - 
\frac{1}{\lambda ({\bf 0})} 
\left( \frac{N_0({\bf 0})}{N_0({\bf x})} -1 \right) \right\}
\] 
and Eq.~(\ref{SV1}) turns out to be just the Schr\"odinger equation for a
3D harmonic oscillator with effective frequency 
\[
\omega_{\Delta} = 4 \pi \sqrt{\frac{2 \lambda+1}{7\lambda \zeta(3)}}
\frac{ k_B T}{m v_F({\bf 0}) R } \equiv \frac{\hbar}{m l_{\Delta}^2} .
\]
The eigenvalues are given by
\begin{equation}
\varepsilon_n =
\frac{2m}{\hbar^2} \frac{\gamma({\bf 0})}{N_0({\bf 0})} 
\left( n+ \frac{3}{2} \right) \hbar \omega_{\Delta}
- \ln{ \left(\frac{T_c({\bf 0})}{T} \right)} . 
\label{eigenv}
\end{equation}

Notice that depending on the temperature, the eigenvalues $\varepsilon_n$
become unstable, i.e., with decreasing temperature they can become negative 
one by one. The temperature
where $\varepsilon_0$ becomes zero, is the critical temperature
$T_c$ of the inhomogeneous system. It is clear from Eq.~(\ref{eigenv})
that due to finite size effects, this critical temperature $T_c$ is {\it lower}
than the homogeneous critical temperature $T_c({\bf 0})$ corresponding to
the density in the center of the trap \cite{baranov}.
Now it is important to realize, that there is always a region sufficiently
close to the critical temperature $T_c$ of the system, where it is
a good approximation to take into account only one single mode 
\begin{equation}
\phi_0({\bf x}) = \left( \frac{1}{\sqrt{\pi} l_{\Delta} } \right)^{3/2}
\exp{\left( -\frac{{\bf x}^2}{2 l_{\Delta}^2} \right) }
\label{psi0}
\end{equation}
in the expansion for $\phi$ and $\xi$ because
the higher modes $\phi_n$ remain stable and their influence thus remains small. 
Furthermore, as will be shown explicitly by a numerical example later on, 
the coupling between modes in the nonlinear terms of Eqs.~ (\ref{actie2FT})
and (\ref{LRA}) 
can be neglected sufficiently close to the critical temperature. 
Plugging everything into the action ${\cal A}$ in Eq.~(\ref{actie2FT}), 
we can first
perform the integral over ${\bf x}$, and subsequently integrate out the
fluctuations that are now solely
represented by the fields $b_0(t)$ and $b_0^{\ast}(t)$.
From the resulting action the effective hamiltonian
for the classical fields $a_0(t)$ and $a_0^{\ast}(t)$ can be obtained 
by introducing their conjugated momenta in the standard way. We find
\begin{eqnarray*}
H_{\Delta} & = &
\frac{8 \imath k_B T}{\pi N_0({\bf 0})} \left\{ 2k_B T \frac{\partial^2}{
\partial a_0 \partial a_0^{\ast} } + 
\frac{\partial}{\partial a_0} [N_0({\bf 0}) \epsilon_0 \right. 
\\ & & \left.  + \beta' |a_0|^2 ] a_0 + \frac{\partial}{\partial a_0^{\ast}}
[N_0({\bf 0}) \epsilon_0 + \beta'|a_0|^2] a_0^{\ast} \right\},
\end{eqnarray*} 
where $\beta' = \beta  \int d{\bf x}\ |\phi_0({\bf x})|^4 =
\beta/(2 \pi l_{\Delta}^2)^{3/2}$.

The equation of motion, or Fokker-Planck equation, 
for the probability distribution $P$ is now given by Schr\"odinger's equation 
\begin{equation}
\imath \hbar \frac{\partial}{\partial t} P = H_{\Delta} P 
\label{SV}
\end{equation}
which has an equilibrium solution 
\[
P(a_0,a_0^{\ast}) \propto \exp{ \left( -\frac{V(a_0,a_0^{\ast})}{k_B T} \right)}
\]
where the potential $V(a_0, a_0^{\ast}) =  N_0({\bf 0}) \epsilon_0
|a_0|^2 + 1/2 \; \beta' |a_0|^4 $. Since $\beta' > 0$, 
this potential has the shape of a parabolic well for $T>T_c$ where
$\varepsilon_0 > 0$,
whereas it becomes mexican-hat shaped below the critical temperature
where $\varepsilon_0 < 0$.

The dynamics of the phase transition can most easily be extracted
if we introduce dimensionless variables $I$, $\theta$, $\tau$, 
and the constant $a$ according to 
\begin{eqnarray} 
a_0 & = &\left( \frac{k_B T}{2 \beta'} \right)^{\frac{1}{4}} 
\sqrt{I} \exp{\imath \theta}
\nonumber \\
t  & = & \frac{\pi N_0({\bf 0}) \hbar}{\sqrt{ 32 (k_B T)^3 \beta'}} \; \tau  
\label{t} \\
a &  = & - \sqrt{\frac{2 \beta'}{k_B T}} \frac{N_0({\bf 0}) \epsilon_0}{\beta'}.
\nonumber
\end{eqnarray}
The Fokker-Planck equation Eq.~(\ref{SV}) can then be rewritten in the form 
\begin{equation}
\frac{\partial}{\partial \tau } P =  \left\{ \frac{\partial^2}{\partial I^2}
4 I + \frac{1}{I} \frac{\partial^2}{\partial \theta^2} +
\frac{\partial}{\partial I} [2(I-a)I -4 ] \right\}P
\label{fpI}
\end{equation}
which exactly is the Fokker-Planck equation associated with the single-mode 
laser \cite{riksen}.
The constant $a$ represents the `pump parameter', and $I$ is equivalent to
the `laser intensity'.
 
The second term on the right-hand side 
of Eq.~(\ref{fpI}) gives rise to phase-diffusion \cite{lew},
but when the system is initially in a state which does not depend on $\theta$,
it will remain phase independent, and we can omit this term from the
Fokker-Planck equation. The stationary solution in terms of the dimensionless
variables becomes
\begin{equation}
P_{\mbox{st}} (I,a) = {\cal N}(a) \exp{ \left( - \frac{(I - a)^2}{4} \right) },
\label{statI}
\end{equation}
where ${\cal N}(a)$ is a normalization factor. This is a Gaussian distribution
centered around $I=a$, where $a<0$ for $T>T_c$, and $a>0$ for $T<T_c$.
Note, however, that $I$ only assumes positive values, so even above the
critical temperature, the average $\langle I \rangle$ is larger than zero.

\begin{figure}[htbp]
\psfig{figure=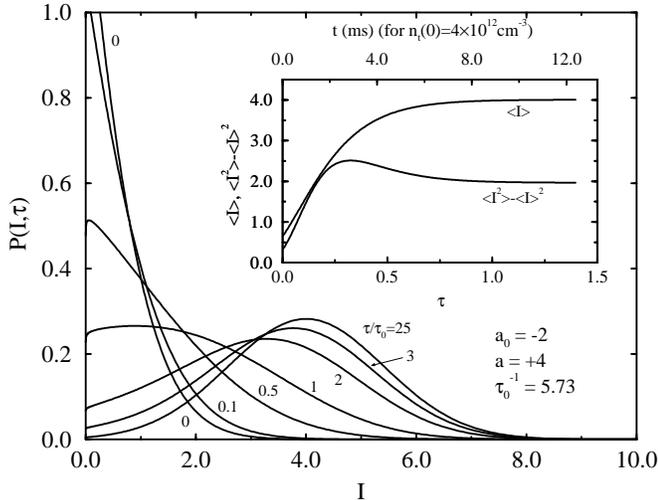}
\caption{\protect\narrowtext
Evolution of the probability distribution for the order parameter in 
normalized units. The inset shows the average value $\langle I \rangle$, 
and its variance as a function of time.}  
\label{fig1}
\end{figure}
One can model the phase transition by changing the constant $a$
suddenly from a negative value $a_0<0$ to a positive value $a>0$ at $\tau=0$. 
Indeed, the process of evaporative cooling is very fast, due to the large
value of the $s$-wave scattering length and corresponding large
cross section for elastic collisions. 
As a consequence, the `translational' degrees of freedom of the atoms 
in the gas are always in thermal equilibrium, but
the order parameter is not and needs more time to adapt. 
The probability distribution 
$P(I,\tau,a)$ will therefore evolve from the stationary solution
$P_{\mbox{st}}(I,a_0)$, see Eq.~(\ref{statI}), at $\tau=0$ to the 
stationary solution $P_{\mbox{st}}(I,a)$ at sufficiently large $\tau$. 
We have solved Eq.~(\ref{fpI}) numerically for $a_0 =-2$ and $a=+4$
using the methods developed in Ref.~\cite{riksen}. The evolution of the
probability distribution $P(I,\tau)$ is plotted for several times
in Fig.~\ref{fig1}. It is clear that after a dimensionless
time of $O(1)$, the
order parameter of the phase transition has reached its stationary
value. This is shown explicitly in the inset of Fig.~\ref{fig1}, where
the average $\langle I\rangle(\tau)$ and the variance $\langle (I -
\langle I\rangle)^2\rangle (\tau)$ are plotted. 

In order to see to what extent the approximations we have made are valid, 
we finish the paper with an explicit numerical example. 
For a gas mixture with density $n_{\alpha}({\bf 0}) = 2\times 10^{12}$
cm$^{-3}$, or equivalently $n_t({\bf 0})= 4\times 10^{12}$ cm$^{-3}$,
one has for the homogeneous critical temperature 
$T_c({\bf 0}) \simeq 37$ nK \cite{us}. If this gas is trapped in a 3D isotropic
harmonic oscillator with $\hbar \omega_0/k_B =7.1$ nK \cite{rice}, 
the extend of the cloud is $R \simeq 11 l$ at $T= 30$ nK, whereas
$l_{\Delta} \simeq 2 l$ is then indeed much smaller. The critical 
temperature $T_c$ according to Eq.~(\ref{eigenv}) is $T_c \simeq 31$ nK.
The second mode of Eq.~(\ref{eigenv}) becomes unstable only at $T \simeq 26$ nK. 
At this temperature the pump parameter $a \simeq 13$, which gives a first
requirement for our theory. However, a more stringent one is that it is only  
a good approximation to take one single mode into account 
if we can also neglect the nonlinear coupling
in the hamiltonian. We must then require that the
interaction $\beta |\phi({\bf 0})|^2$ is small compared to $N_0({\bf 0})
\hbar \omega_{\Delta}$. This is true for $T> 29.3$ nK, from
which it follows that the approximations are valid if $a \leq 4.3$.
For $T=29.3$ nK, the order parameter $\Delta({\bf 0})/ k_BT \simeq 1$.

In Fig.~\ref{fig1} we therefore have chosen $a=4$, and from
the inset of this figure we conclude that for a total density
$n_t({\bf 0})=4 \times 10^{12}$ cm$^{-3}$, one unit of dimensionless
time $\tau$ corresponds to $8.9$ ms, which is much larger than the
$\hbar/ k_B T_c({\bf 0}) =0.2$ ms as estimated in our previous work \cite{us},
because Eq.~(\ref{t}) clearly shows that the time scale depends also on 
several other relevant parameters. In addition, 
this is caused by a critical slowing down, which 
can be reduced by increasing the final value of the pump parameter
$a$, but requires inclusion of more modes in the theory. In a
future publication we will come back to this multi-mode regime.
However, within the present theory,
for a lifetime of the gas of the order of the nucleation time, one needs
a magnetic bias field of about $0.7$ T. A better option is to trap
and cool the lowest two hyperfine states in an optical trap, which has recently
been shown to be possible for bosons \cite{MIT2}. In any case, we
believe that this paper, in combination with Ref.~\cite{us},
shows that there are no theoretical limitations to achieve
a BCS transition in atomic $^6$Li.  


\end{multicols}

\end{document}